\documentclass[pra,twocolumn,amsmath,amssymb,floatfix,reprint,footinbib,superscriptaddress,longbibliography,showkeys]{revtex4-2}
\usepackage{picinpar,graphicx}
\usepackage{braket}
\usepackage{amsmath}
\usepackage{graphicx}
\usepackage{epstopdf}
\usepackage{dcolumn}
\usepackage{graphicx}
\usepackage{mathrsfs}
\usepackage{mdwlist}
\usepackage{subfigure}
\usepackage{booktabs}
\usepackage{amsmath}
\usepackage{dsfont}
\usepackage{amstext}
\usepackage{amssymb}
\usepackage{amsbsy}
\usepackage{bbm}
\usepackage{amsthm}
\usepackage{graphicx}
\usepackage{textcomp}
\usepackage{multirow}
\usepackage[utf8]{inputenc}  
\usepackage[T1]{fontenc}     

\usepackage[colorlinks,citecolor=blue]{hyperref}
\usepackage{url}
\usepackage[colorlinks]{hyperref}
\hypersetup{%
	plainpages=true,
	breaklinks=true,  
	hypertexnames=false,  
	pageanchor=true,
	colorlinks=true,
	linkcolor={blue},
	citecolor={red},
	urlcolor={blue},
	anchorcolor={black}
}

\begin{document}
\title{Entanglement protection induced by mixed noise}
\author{Tengtao Guo}
\email{These authors contributed equally to this work.}

\author{Yuxuan Zhou}
\email{These authors contributed equally to this work.}

\author{Jiahui Feng}
\author{Xinyu Zhao}
\email{xzhao@fzu.edu.cn}

\author{Yan Xia}
\email{xia-208@163.com}

\affiliation{Fujian Key Laboratory of Quantum Information and Quantum Optics, Fuzhou
University, Fuzhou 350116, China}
\affiliation{Department of Physics, Fuzhou University, Fuzhou 350116, China}
\begin{abstract}
Contrary to the conventional view that noise is detrimental, we show
that mixed noise can protect entanglement in a two-atom-cavity system.
Specifically, the leakage of the cavity and the stochastic atom-cavity
couplings are modeled as two types of noises. From the analytical
derivation of the dynamical equations, the mechanism of the entanglement
protection is revealed as the high-frequency (HF) noise in the atom-cavity
couplings could suppress the decoherence caused by the cavity leakage,
thus protect the entanglement. We investigate the entanglement protection induced by mixed noise constructed from diverse noise types, including the Ornstein-Uhlenbeck noise, flicker noise, and telegraph noise. Numerical simulations demonstrate that
entanglement protection depends critically on the proportion of HF components in the power spectral density of the mixed noise.
Our work establishes that enhanced HF components are essential for effective noise-assisted entanglement
protection, offering key insights for noise engineering in practical
open quantum systems.
\end{abstract}
\maketitle

\section{INTRODUCTION}

The concept of quantum entanglement originated from debates in the
early development of quantum mechanics \cite{Einstein1935PR,Bohr1935PR}.
As a distinctive feature of the quantum world, entanglement has attracted
considerable attention due to its profound theoretical significance
and practical applications \cite{Su2020PRA,Gu2025AQT,Kang2023OE,Liu2023PRA,Qiu2025NJP,Zheng2020PRA,Zheng2019AP}.
Theoretically, studies of quantum entanglement have advanced our understanding
of microscopic physical phenomena. Practically, quantum entanglement
has become an essential resource for various quantum information processing
protocols, including super dense coding \cite{Muralidharan2008PRA},
quantum key distribution \cite{Cao2020PRL,Harn2020IC,Scarani2009RMP},
quantum teleportation \cite{Bouwmeester1997N,Barrett2004N,Bandyopadhyay2006PRA,Pirandola2015NP,Ren2017N,Zeilinger2018NP},
and quantum networks \cite{Cirac1997PRL,Kang2017APa,Lu2010IJQI,Wang2025APL,Stolk2022PQ}.
Particularly, with the demonstration of quantum computational advantages
in experiments, quantum entanglement has emerged as a key to realizing
universal quantum computing \cite{Blais2004PRA,Burkard2020NRP,Duan1998PRA,Arute2019N,Zhong2020S}.
However, entanglement is extremely fragile when quantum systems are
subjected to environmental noise \cite{Xiang2025PRA,Zhao2019OEa,Zhao2025OE}.
Therefore, protecting entanglement against noise-induced decoherence
is of critical importance \cite{Zurek2003RMP,Bellomo2007PRL,Chen2019PRA,Yu2009S,Zhou2021PRA,Yang2022OE,Xiong2019OE}.

The vast majority of existing schemes for protecting quantum entanglement
aim to mitigate the impact of noise \cite{Burgelman2025PQ,Chen2018PRAa,Kang2023NJP,Ni2023N,Piccolini2023AQT,Wang2025APL}.
For instance, dynamical decoupling schemes \cite{Viola1999PRL,Uhrig2008NJP,Burgelman2025PQ}
seek to eliminate the effects of noise through periodic driving pulses.
Decoherence-free subspace schemes \cite{Lidar1998PRL,Wu2017AP}, on
the other hand, strive to identify subspace that remain unaffected
by noise. Meanwhile, quantum feedback control schemes \cite{Ahn2002PRA,Amazioug2025EPJP,Song2012JOSAB,Song2012PRA,Yang2022OE}
aim to compensate for the impact of noise via feedback mechanisms.
However, noise is ubiquitous and inevitable to some extent. A compelling
question arises: Could noise itself be harnessed as a resource? If
achievable, employing noise (as opposed to artificially engineered
operations) to mitigate the deleterious effects of other noise sources
may constitute an elegant strategy for quantum entanglement protection.

In Refs. \cite{Jing2013SR,Jing2018PRA,Zhao2022PRA}, the authors have
presented several successful cases where quantum entanglement is protected
by utilizing noise. Nevertheless, there remain numerous open questions
in this research field. For instance, noise typically originates from
multiple sources, meaning real-world noise is often a mixture of several
types. How the mixed noise affects entanglement protection constitutes
one of the key open questions. In this paper, we will focus on exploring
this specific issue. 

\begin{figure}
\includegraphics[width=1\columnwidth]{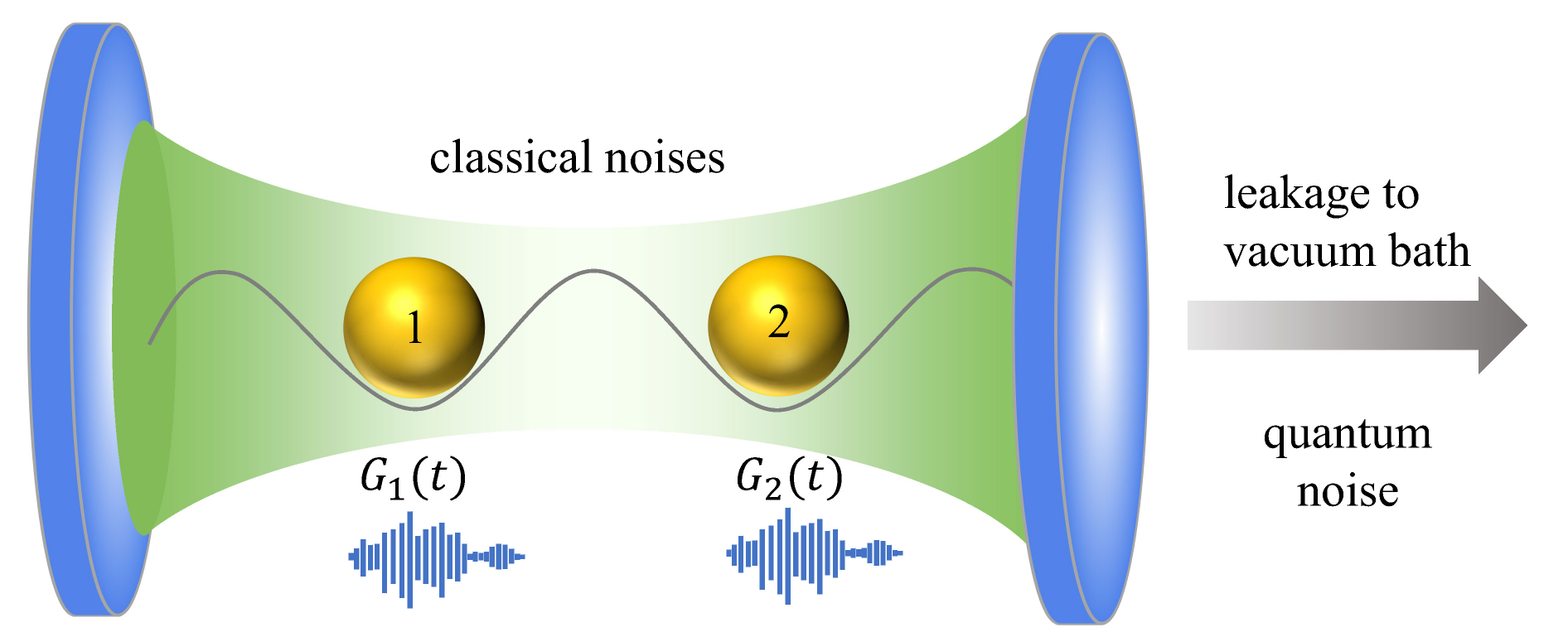} 

\caption{Schematic diagram of two-level atoms interacting with an optical cavity.
The cavity leaks into a vacuum bath. The coupling strengths between
each atom and the cavity depends on the positions of the atoms within
the electromagnetic field. Random motion of atoms induces stochastic
coupling strengths $G_{1}(t)$ and $G_{2}(t)$.}
\label{fig1}
\end{figure}

To be specific, we consider two atoms inside a cavity as shown in
Fig.~\ref{fig1}. By introducing stochastic atom-cavity couplings
$G_{i}(t)$ (modeled as classical noises), the entanglement loss caused
by the leakage of the cavity (modeled as quantum noise) can be suppressed.
From the analytical derivation of the dynamical equations, the mechanism
of entanglement protection is revealed as rapidly varying noise in
the atom-cavity couplings could freeze the quantum state to its initial
state, thus protect entanglement. Then, we use numerical simulations
to evaluate the performance of entanglement protection for several
types of mixed classical noise. The properties of each component of
the mixed noise is proved to be crucial to the performance. The performance
of entanglement protection by mixed classical noise may be superior
to that of any individual noise, or it may also be worse than that
of individual noises. All the numerical results can be explained by
using the physical picture ``high-frequency (HF) classical noises
can suppress the decoherence caused by low-frequency (LF) quantum
noise''. This provides a guideline for how to select the properties
of various components in mixed noise to achieve a better performance
of entanglement protection.

The remainder of this paper is organized as follows: In Sec.~\ref{sec:2}~,
we describe the theoretical model and the sources of noises. In Sec.~\ref{sec:3},
the dynamical equation of the system is derived and the mechanism
of the entanglement protection is revealed. In Sec.~\ref{sec:4},
we numerically evaluate the performance of entanglement protection
for the properties of mixed noises. Finally, the conclusion is presented
in Sec.~\ref{sec:5}

\section{\label{sec:2}Atom-cavity system with two types of noises}

As shown in Fig.~\ref{fig1}, we consider two atoms interacting with
a single mode cavity, the Hamiltonian can be written as \cite{Jaynes1963PI}
\begin{align}
H_{{\rm tot}} & =\sum_{i=1}^{2}\left\{ \frac{\omega}{2}\sigma_{z}^{(i)}+G_{i}(t)\left[a\sigma_{+}^{(i)}+a^{\dag}\sigma_{-}^{(i)}\right]\right\} +\varOmega a^{\dag}a\nonumber \\
 & +\sum_{k}\omega_{k}b_{k}^{\dag}b_{k}+g_{k}a^{\dag}b_{k}+g_{k}^{*}ab_{k}^{\dag},\label{eq:Htot}
\end{align}
where $\frac{\omega}{2}\sigma_{z}^{(i)}$ is the Hamiltonian of two
atoms, $\varOmega a^{\dag}a$ is the Hamiltonian of the cavity, and
$G_{i}(t)\left[a\sigma_{+}^{(i)}+a^{\dag}\sigma_{-}^{(i)}\right]$
are the interactions between the atoms and the cavity. The time-dependent
coupling strengths are described by $G_{i}(t)$ ($i=1,2$). The cavity
is leaking to a vacuum environment described by $H_{{\rm E}}=\sum_{k}\omega_{k}b_{k}^{\dag}b_{k}$,
where $b_{k}^{\dag}$ and $b_{k}$ are the creation and annihilation
operators for mode $k$. Assuming $\omega=\Omega$, the Hamiltonian
in the interaction picture yields 
\begin{equation}
H_{{\rm I}}=\sum_{i=1}^{2}G_{i}(t)\left[a\sigma_{+}^{(i)}+a^{\dag}\sigma_{-}^{(i)}\right]+\sum_{k}\left(g_{k}a^{\dag}b_{k}e^{-i\Delta_{k}t}+{\rm h.c.}\right).\label{eq:HI1}
\end{equation}
If we define a collective dissipation operator $B(t)\equiv\sum_{k}g_{k}b_{k}e^{-i\Delta_{k}t}$,
the Hamiltonian can be also written as 
\begin{equation}
H_{{\rm I}}=\sum_{i=1}^{2}G_{i}(t)\left[a\sigma_{+}^{(i)}+a^{\dag}\sigma_{-}^{(i)}\right]+\left[a^{\dag}B(t)+{\rm h.c.}\right].\label{eq:HI2}
\end{equation}

There are two types of time dependent terms in the Hamiltonian~(\ref{eq:HI2}),
corresponding to two types of noises, quantum noise and classical
noise. For $B(t)\equiv\sum_{k}g_{k}b_{k}e^{-i\Delta_{k}t}$, it is
an operator, the complexity of its evolution originates from the huge
degrees of freedom of the environment described by the quantized Hamiltonian
$H_{{\rm E}}=\sum_{k}\omega_{k}b_{k}^{\dag}b_{k}$. Since the environment
contains infinite modes, the complicated impact from the environment
can be regarded as a stochastic process, reflected by the operator
$B(t)$. Here, the evolution of $B(t)$ is governed by quantum mechanical
equations. In this sense, the influence originating from the operator
$B(t)$ can be regarded as a type of quantum noise.

Besides the quantum noise, there are also other time-dependent terms
$G_{i}(t)$ in Eq.~(\ref{eq:HI2}). The time dependent coupling strengths
originate from the dipole interactions between the atoms and the cavity
field. The electromagnetic field typically forms a standing wave inside
the cavity (see Fig.~\ref{fig1}), so the dipole coupling is sensitive
to the position of the atoms \cite{Rempe1987PRL}. When the atoms
are located at the node (antinode) of the standing wave, the coupling
strengths are the minimum (maximum). According to Refs.~\cite{Wu1997PRL,Mabuchi1999APBLO,Tian2008NJP},
the coupling can be expressed as
\begin{equation}
G_{i}(t)=G_{0i}\sin\left\{ \kappa\left[x_{0i}+\xi_{i}(t)\right]\right\} ,\quad(i=1,2)
\end{equation}
where $\kappa$ is the wave number, $x_{0i}$ is the balanced position
and $\xi_{i}(t)$ is a stochastic function (noise) describing a random
deviation from the balanced position. Throughout the paper, we always
assume $\xi_{1}(t)=\xi_{2}(t)=\xi(t)$, so we use a single notation
$\xi(t)$ for both classical noises.

The properties of noises are mainly characterized by the correlation
functions. For quantum noise $B(t)$ which is described by quantized
operators, the correlation function is defined as
\begin{equation}
K_{{\rm Q}}(t,s)=\langle B(t)B^{\dag}(s)\rangle=\sum_{k}|g_{k}|^{2}e^{-i\omega_{k}(t-s)},\label{eq:KQ}
\end{equation}
where $\langle\cdot\rangle$ represents the mean value in quantum
mechanics. In the limit of infinite modes, the summation in Eq.~(\ref{eq:KQ})
can be replaced by an integration as $\sum_{k}|g_{k}|^{2}e^{-i\omega_{k}(t-s)}\rightarrow\int_{0}^{\infty}d\omega|g(\omega)|^{2}e^{-i\omega(t-s)}$.
If we further defined the power spectrum density (PSD) as $J(\omega)=|g(\omega)|^{2}$,
the correlation function can be regarded as a Fourier transformation
of the PSD $K_{{\rm Q}}(t,s)=\int_{0}^{\infty}d\omega J_{{\rm Q}}(\omega)e^{-i\omega(t-s)}$.

In the following sections, we will use the Lorentzian PSD for quantum
noise $B(t)$ as
\begin{equation}
J_{{\rm Q}}(\omega)=\frac{1}{2\pi}\frac{\Gamma_{{\rm Q}}\gamma_{{\rm Q}}^{2}}{\omega^{2}+\gamma_{{\rm Q}}^{2}},\label{eq:JQ}
\end{equation}
where $\Gamma_{{\rm Q}}$ is a constant global dissipation rate. It
is straightforward to calculate the corresponding correlation function
for this Lorentzian PSD as
\begin{equation}
K_{{\rm Q}}(t,s)=\int_{0}^{\infty}d\omega J_{{\rm Q}}(\omega)e^{-i\omega(t-s)}=\frac{\Gamma_{{\rm Q}}\gamma_{{\rm Q}}}{2}e^{-\gamma_{{\rm Q}}|t-s|}.\label{eq:KQ2}
\end{equation}
This is called the Ornstein-Uhlenbeck (O-U) correlation function \cite{Tu2008PRB,Breuer2016RMP}.
From Eq.~(\ref{eq:KQ2}), the parameter $\gamma_{{\rm Q}}$ has the
dimension of the reciprocal of time. So, it is clear that $\tau_{{\rm Q}}=1/\gamma_{{\rm Q}}$
has the dimension of time and it can be used to characterize the correlation
time. When $\tau_{{\rm Q}}\rightarrow0$, $\gamma_{{\rm Q}}\rightarrow\infty$,
$K_{{\rm Q}}$ becomes a delta function, indicating no correlations
for two different time points $t$ and $s$. In this case, the noise
$B(t)$ is considered as a Markovian noise. When $\tau_{{\rm Q}}$
is finite, two different time points $t$ and $s$ will have non-zero
correlation, the noise $B(t)$ is then considered as non-Markovian
noise.

For classical noises, the correlation function is defined as
\begin{equation}
K_{\xi}(t,s)=\mathcal{M}[\xi(t),\xi(s)]
\end{equation}
where $\mathcal{M}[\cdot]$ denotes the statistical average over stochastic
variables. Similarly, $K_{\xi}$ can be also expressed as a Fourier
transformation $K_{\xi}(t,s)=\int_{0}^{\infty}d\omega J_{\xi}(\omega)e^{-i\omega(t-s)}$.
The PSD can be defined in a similar way. In the following discussion,
we will consider three types of classical noises and their mixture.

(a) O-U noise with the corresponding PSD
\begin{equation}
J_{{\rm \xi}}(\omega)=\frac{1}{2\pi}\frac{\Gamma_{{\rm \xi}}\gamma_{{\rm \xi}}^{2}}{\omega^{2}+\gamma_{{\rm \xi}}^{2}},\label{eq:JOU}
\end{equation}
which is similar to Eq.~(\ref{eq:JQ}) except the sub-index. The
key parameter is also $\gamma_{\xi}$, where $1/\gamma_{\xi}$ determines
the correlation time.

(b) Flicker noise with the corresponding PSD
\begin{equation}
J_{{\rm F}}(\omega)=A\omega^{\eta}.\label{eq:JF}
\end{equation}
The key parameter in the flicker noise is $\eta$. When $\eta=-1$,
$J_{{\rm F}}$ is inverse proportional to the frequency $\omega$,
this is called 1/f noise (or pink noise). Charge noise in semiconductors
is often in this form \cite{Zajac2018S,Zhao2022PRAa,Culcer2009APL,Yoneda2018NN}.
Besides $\eta=-1,$ when $\eta=-2$, $\eta=1$, and $\eta=2$, the
corresponding noises are called ``red'', ``blue'', and ``violet''
noises.

(c) Telegraph noise that is generated by the following process. A
telegraph noise function $\xi(t)$ has only two possible values $\xi(t)=\pm1$.
This type of noise is generated by separating the evolution into $N$
steps. At each step, the stochastic variable $\xi(t)$ changes its
sign as $\xi(t_{i+1})=-\xi(t_{i})$ with a probability $p_{{\rm jump}}$.
Correspondingly, there is also a probability of $1-p_{{\rm jump}}$
that $\xi(t_{i+1})=\xi(t_{i})$ keeps unchanged \cite{Cheng2008PRA,Culcer2009APL,Zhao2022PRA}. 

In Sec.~\ref{sec:4}, we will show the influence of the mixture of
these noises.

\section{\label{sec:3}Mechanism of entanglement protection}

In the atom-cavity system we have discussed in Sec.~\ref{sec:2},
there are two sources of noise. One is the leakage of the cavity,
which has been modeled as a quantum noise $B(t)$, the other is the
stochastic positions of two atoms, which is modeled as a classical
noise $\xi(t)$. In this section, we will show how the classical noise
could suppress the decoherence caused by the quantum noise and thus
protect entanglement.

The dynamics of whole system including the environment is governed
by the Schr\"{o}dinger equation:dynamical decoupling
\begin{equation}
\frac{\partial}{\partial t}|\psi(t)\rangle=-iH_{{\rm I}}|\psi(t)\rangle,\label{eq:SEQ}
\end{equation}

\noindent where the interaction Hamiltonian is given in Eq.~(\ref{eq:HI1}).
When there is only one excitation in the initial state, the dynamical
evolution of the system will be confined to a Hilbert subspace spanned
by the following basis vectors: $|e,g,0,0_{k}\rangle$, $|g,e,0,0_{k}\rangle$,
$|g,g,1,0_{k}\rangle$, $|g,g,0,1_{k}\rangle$, and $|g,g,0,0_{k}\rangle$,
where $|g\rangle$ and $|e\rangle$ represent the ground and excited
state of two atoms, $|0\rangle$ and $|1\rangle$ represent the vacuum
and first excited state of the cavity, $|0_{k}\rangle$ and $|1_{k}\rangle$
represent the vacuum and first excited state of the $k^{{\rm th}}$
mode of the environment. Then, an arbitrary state can be written as
a linear combination of these basis as
\begin{flalign}
|\psi(t)\rangle= & C_{1}(t)|e,g,0,0_{k}\rangle+C_{2}(t)|g,e,0,0_{k}\rangle\nonumber \\
 & +C_{3}(t)|g,g,1,0_{k}\rangle+D_{k}(t)|g,g,0,1_{k}\rangle\nonumber \\
 & +C_{4}(t)|g,g,0,0_{k}\rangle,\label{eq:psitot}
\end{flalign}
Substituting Eq.~(\ref{eq:psitot}) into Eq.~(\ref{eq:SEQ}), one
can obtain the dynamical equations for the probability amplitudes
$C_{i}(t)$ as
\begin{equation}
\frac{d}{dt}C_{1}(t)=-iG_{1}(t)C_{3}(t),\label{eq:dC1}
\end{equation}
\begin{equation}
\frac{d}{dt}C_{2}(t)=-iG_{2}(t)C_{3}(t),\label{eq:dC2}
\end{equation}
\begin{flalign}
\frac{d}{dt}C_{3}(t) & =-iG_{1}(t)C_{1}(t)-iG_{2}(t)C_{2}(t)\nonumber \\
 & -i\sum_{k}g_{k}e^{-i\Delta_{k}t}D_{k}(t),\label{eq:dC3}
\end{flalign}
\begin{equation}
\frac{d}{dt}D_{k}(t)=-ig_{k}^{*}e^{i\Delta_{k}t}C_{3}(t),\label{eq:dDk}
\end{equation}
\begin{equation}
\frac{d}{dt}C_{4}(t)=0,
\end{equation}
For a initial vacuum environment, we have boundary conditions $D_{k}(0)=0$.
Integrating Eq.~(\ref{eq:dDk}), we have

\begin{flalign}
D_{k}(t) & =D_{k}(0)+\int_{0}^{t}-ig_{k}^{*}e^{i\omega_{k}s}C_{3}(s)ds\nonumber \\
 & =\int_{0}^{t}-ig_{k}^{*}e^{i\omega_{k}s}C_{3}(s)ds.\label{eq:Dk}
\end{flalign}
Then, we can substitute Eq.~(\ref{eq:Dk}) into Eq.~(\ref{eq:dC3}),

\begin{flalign}
\frac{d}{dt}C_{3}(t)= & -iG_{1}(t)C_{1}(t)-iG_{2}(t)C_{2}(t)\nonumber \\
 & -\int_{0}^{t}\sum_{k}|g_{k}|^{2}e^{-i\omega_{k}(t-s)}C_{3}(s)ds.\label{eq:dC32}
\end{flalign}
The term $\sum_{k}|g_{k}|^{2}e^{-i\omega_{k}(t-s)}=K_{{\rm Q}}(t,s)$
is just the correlation function of the environment. As discussed
in Sec.~\ref{sec:2}, we assume the PSD of the quantum noise satisfy
the Lorentzian form with the corresponding correlation function shown
in Eq.~(\ref{eq:KQ}). Then, one can define the integration in the
right-hand-side of Eq.~(\ref{eq:dC32}) as $I(t)\equiv\int_{0}^{t}\frac{\Gamma_{{\rm Q}}\gamma_{{\rm Q}}}{2}e^{-\gamma_{{\rm Q}}|t-s|}C_{3}(s)ds$.
It is straightforward to derive the differential equation for $I(t)$
as

\begin{equation}
\frac{d}{dt}I(t)=-\gamma_{{\rm Q}}I(t)+\frac{\Gamma_{{\rm Q}}\gamma_{{\rm Q}}}{2}C_{3}(t).\label{eq:dI}
\end{equation}
By solving Eqs.~(\ref{eq:dC1},\ref{eq:dC2},\ref{eq:dC32},\ref{eq:dI}),
one can obtain the solutions of the probability amplitudes $C_{i}(t)$
($i=1,2,3$). The solution for $C_{4}(t)$ is obviously $C_{4}(t)=C_{4}(0)$.
Thus, the time evolution of the two atoms is completely determined.

Here, we study the initial state $|\psi(0)\rangle=\frac{1}{\sqrt{2}}(|e,g\rangle+|g,e\rangle)\otimes|0,0_{k}\rangle$,
namely two atoms are prepared in the maximally entangled Bell state
and both the cavity and the environment are in vacuum states. Then,
we will focus on the dynamics of the entanglement of the two atoms,
which can be measured by ``concurrence'' proposed in Ref.~\cite{Wootters1998PRL}.
The solution for Eq.~(\ref{eq:dC1}) and Eq.~(\ref{eq:dC2}) can
be formally written as
\begin{equation}
C_{i}(t)=C_{i}(0)-i\int_{0}^{t}G_{i}(s)C_{3}(s)ds.\quad(i=1,2)\label{eq:Ca}
\end{equation}
When $G_{i}(t)$ contains a HF noise $\xi(t)$, in a short period
$s\in[t_{1},t_{2}]$, $C_{3}(s)$ can be regarded as a constant if
it is varying much slower than $G_{i}(t)$. This leads to $i\int_{t_{1}}^{t_{2}}G_{i}(s)C_{3}(s)ds=iC_{3}(t)\int_{t_{1}}^{t_{2}}G_{i}(s)ds=0$.
Because $G_{i}(t)$ is a zero-mean stochastic function, its integration
over a short period is zero. Finally, we obtain $C_{i}(t)=C_{i}(0)$
meaning that $C_{1}(t)$ and $C_{2}(t)$ do not change in the time
evolution, and the quantum state remains the initial state. This is
the mechanism of entanglement protection.

In a word, a HF noise can freeze the initial quantum state and protect
entanglement. Actually, some other entanglement protection schemes
such as dynamical decoupling \cite{Viola1999PRL,Khodjasteh2005PRL}
and quantum Zeno effect\cite{Misra1977JMP,Ai2010PRA,Liu2021AP,Wu2020OE}
are based on similar mathematical mechanism. With the conclusion that
entanglement protection prefers HF noise, we will study the mixture
of several widely studied noises and try to explore the impact of
various parameters on quantum entanglement protection.

\section{\label{sec:4}Entanglement protection by mixed noise}

The mechanism of entanglement protection has been revealed in Sec.~\ref{sec:3},
and examples of using HF noises to protect quantum coherence (although
not entanglement) have been shown in Ref.~\cite{Zhao2022PRA}. There
is still an open question. In natural world, noises typically originate
from multiple sources and noises are difficult to be modeled as being
characterized by a single type of PSD. Thus, we will investigate the
impact of the mixture of two types of classical noises $\xi_{a}(t)$
and $\xi_{b}(t)$, which can be written as 
\begin{equation}
\xi(t)=p\xi_{a}(t)+(1-p)\xi_{b}(t),\quad(p\in[0,1])\label{eq:mixture}
\end{equation}
where $p$ indicates the mixing ratio of two noises. Both $\xi_{a}(t)$
and $\xi_{b}(t)$ can be O-U noise, flicker noises, or telegraph noise.
In the following subsections, we will show the mixture is significant
for entanglement protection and all the numerical results will be
explained by the mechanism revealed in Eq.~(\ref{eq:Ca}).

\subsection{\label{sec:4A}The Influence of the Properties of Classical Noise}

\begin{figure}[h]
\includegraphics[width=1\columnwidth]{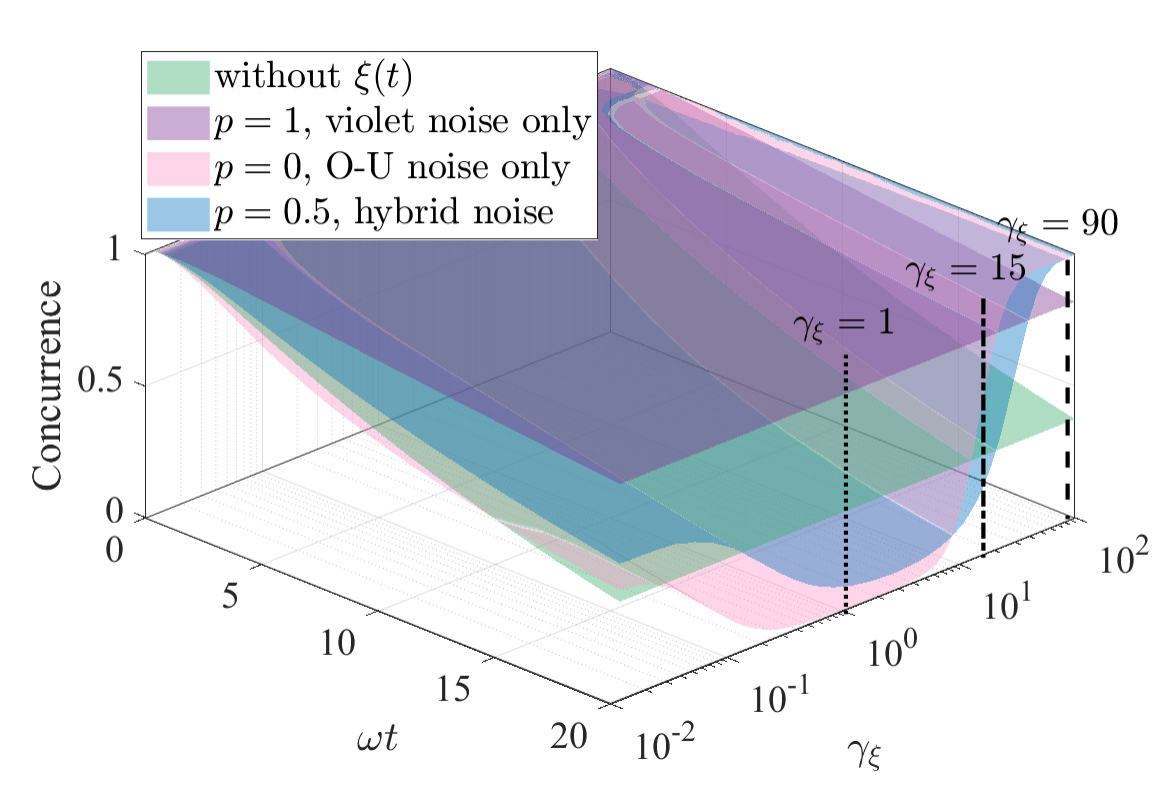}

\caption{\label{fig:2}Quantum entanglement protection induced by the mixture
of violet noise ($\xi_{a}$) and O-U noise ($\xi_{b}$). The violet,
pink, and blue surfaces indicate the time evolution of concurrence
under different mixture of violet and O-U noise. The green surface
is presented as a comparison to demonstrate the case without noise. }
\end{figure}
\begin{figure}
\includegraphics[width=1\columnwidth]{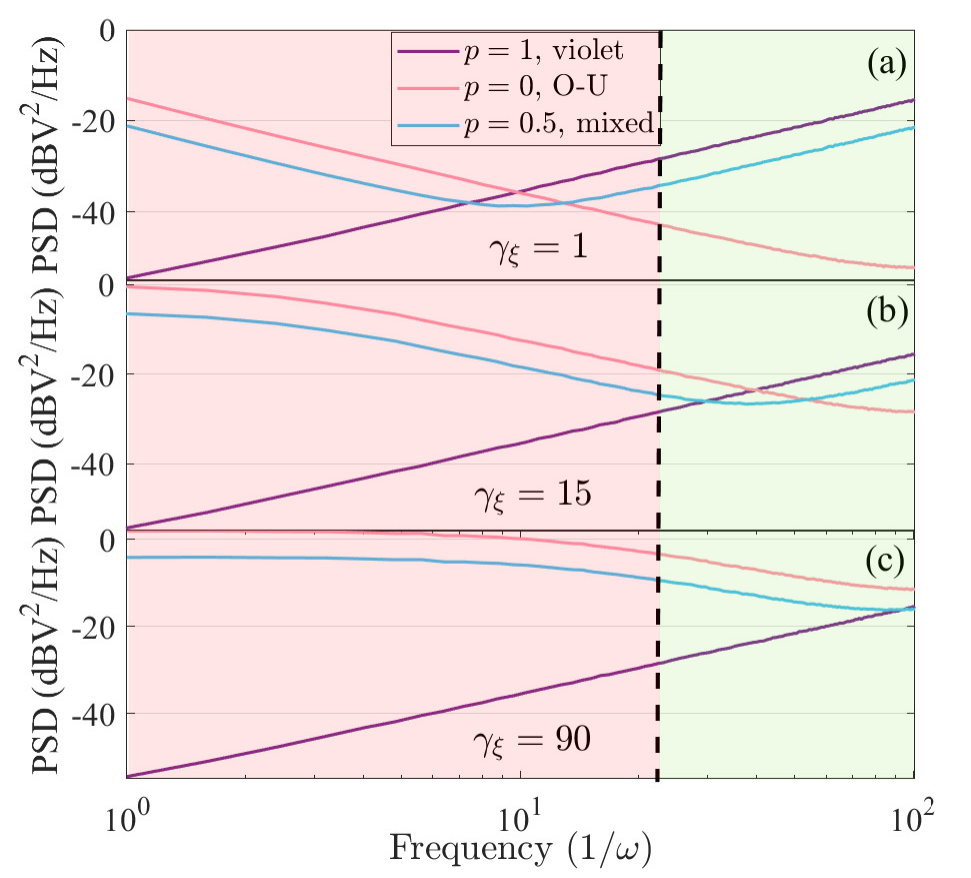}

\caption{\label{fig:3}PSD for three special cases $\gamma_{\xi}=1$, $\gamma_{\xi}=15$,
and $\gamma_{\xi}=90$. The PSD for violet noise ($p=1$), O-U noise
($p=0$), and mixed noise ($p=0.5$) are represented by the purple,
pink, and blue curves. The black dashed line serves as a phenomenological,
approximate boundary to delineate the LF and HF regimes, reflecting
a heuristic distinction rather than a rigid, physically inherent threshold.}
\end{figure}

We start from the mixture of violet noise and O-U noise. The numerical
results are presented in Fig.~\ref{fig:2}. We focus on the impact
of the memory time indicated by the parameter $\tau_{\xi}=1/\gamma_{\xi}$
as shown in Eq.~(\ref{eq:JOU}). Since the time evolution represented
by the violet and green surfaces do not involve O-U noise, these two
surfaces are not affected by the parameter $\gamma_{\xi}$. As a comparison,
the pink and blue surfaces are affected by the parameter $\gamma_{\xi}$
because they contain O-U noise either directly or indirectly.

For the pink surface with O-U noise only ($p=0$), large $\gamma_{\xi}$
(e.g., $\gamma_{\xi}=100$) leads to a better performance of entanglement
protection. This is consistent with the entanglement protection mechanism
presented in the Sec.~\ref{sec:3}. The PSD in Eq.~(\ref{eq:JOU})
contains more HF component when $\gamma_{\xi}$ is large. Since entanglement
protection requires HF noise, it is straightforward that $\gamma_{\xi}=100$
lead to a better performance. When $\gamma_{\xi}$ decreases, the
residue entanglement also decreases. At $\gamma_{\xi}=1$, the residue
entanglement is even smaller than the case without noise, since it
contains more LF noise and fewer HF noise. Certainly, when $\gamma_{\xi}\rightarrow0$,
$J_{\xi}(\omega)\rightarrow0$, the residue entanglement eventually
converges to the case without noise (green surface).

For the blue surface with equally mixed noise ($p=0.5$), the shape
of the surface basically follows the similar pattern as the pink one
since it contains the O-U noise. However, it is worth to note that
the order of entanglement protection performance for the three cases
($p=0$, $p=0.5$, $p=1$) could be different when $\gamma_{\xi}$
changes. The performance of entanglement protection for the mixed
noise is not always intermediate between that of the two types of
noises only. The influence of the parameter $\gamma_{\xi}$ could
be complicated. 

Although the numerical conclusions presented in Fig.~\ref{fig:2}
may seem highly complicated, we will demonstrate that all these phenomena
can be explained by the mechanism revealed in Eq.~(\ref{eq:Ca}),
namely, the quantum noise from the cavity leakage can be suppressed
by a HF noise. We illustrate this mechanism with three representative
values of $\gamma_{\xi}$ ($\gamma_{\xi}=1$, $\gamma_{\xi}=15$,
and $\gamma_{\xi}=90$), which are represented by dashed lines, dash-dot
lines, and dotted lines in Fig.~\ref{fig:2}, respectively. 

For $\gamma_{\xi}=90$, the mixed noise performs better than the violet
noise but worse than the O-U noise. In Fig.~\ref{fig:3}(a), we plot
the PSD for three noises. To suppress quantum noise, the so-called
``HF'' noise mentioned herein refers to the frequencies higher than
that of quantum noise. In numerical simulations, quantum noise encompasses
all frequency components, and its distribution is determined by $\gamma_{{\rm Q}}$.
Here, we phenomenologically use a dashed line in Fig.~\ref{fig:3}
to approximately represent the typical frequency of quantum noise:
the region to the right of this dashed line (light green area) can
be considered as the frequency range higher than that of quantum noise,
while the region to the left (pink area) corresponds to the frequency
range lower than that of quantum noise. In Fig.~\ref{fig:3}(a),
it is evident that among the HF components (light-green region), the
violet noise contains the smallest proportion of HF noise (roughly
measured by the area under the curve), while O-U noise contains the
highest proportion of HF noise, and the proportion of HF noise in
mixed noise falls between the two types of noise. Consequently, in
Fig.~\ref{fig:2}, when $\gamma_{\xi}=90$, the O-U noise achieves
the best protection effect, the mixed noise provides slightly less
protection, and the violet noise has the worst protection effect.

For $\gamma_{\xi}=15$, as shown in Fig.~\ref{fig:3}(b), in the
HF region (light-green region), the proportion of HF noise in mixed
noise is the smallest, while the proportion of HF noise in O-U noise
is higher than in mixed noise but lower than in violet noise. Therefore,
in Fig.~\ref{fig:2}, when $\gamma_{\xi}=15$, the order of protection
performance from best to worst is purple noise, O-U noise, and mixed
noise.

For $\gamma_{\xi}=1$, the mixed noise performs better than the O-U
noise but worse than the violet noise. It can be observed from Fig.~\ref{fig:3}(c)
that in the HF region (light green region), the proportion of HF noise
in violet noise is the highest, which results in the best entanglement
protection effect in Fig.~\ref{fig:2}. In contrast, O-U noise has
the lowest proportion of HF noise in the HF region, leading to the
weakest protection performance. The proportion of HF noise in mixed
noise falls between O-U noise and violet noise, and so does its protection
performance.

In conclusion, the above analysis reaffirms the entanglement protection
mechanism discussed in Sec.~\ref{sec:3}, namely that the (relative)
HF classical noise can suppress the dissipative effects induced by
the (relative) LF quantum noise, thereby preserving quantum entanglement.
All numerical results presented in Fig.~\ref{fig:2} can be perfectly
explained by this mechanism reflected in Eq.~(\ref{eq:Ca}).

\subsection{The Influence of the Properties of Quantum Noise}

\begin{figure}
\includegraphics[width=1\columnwidth]{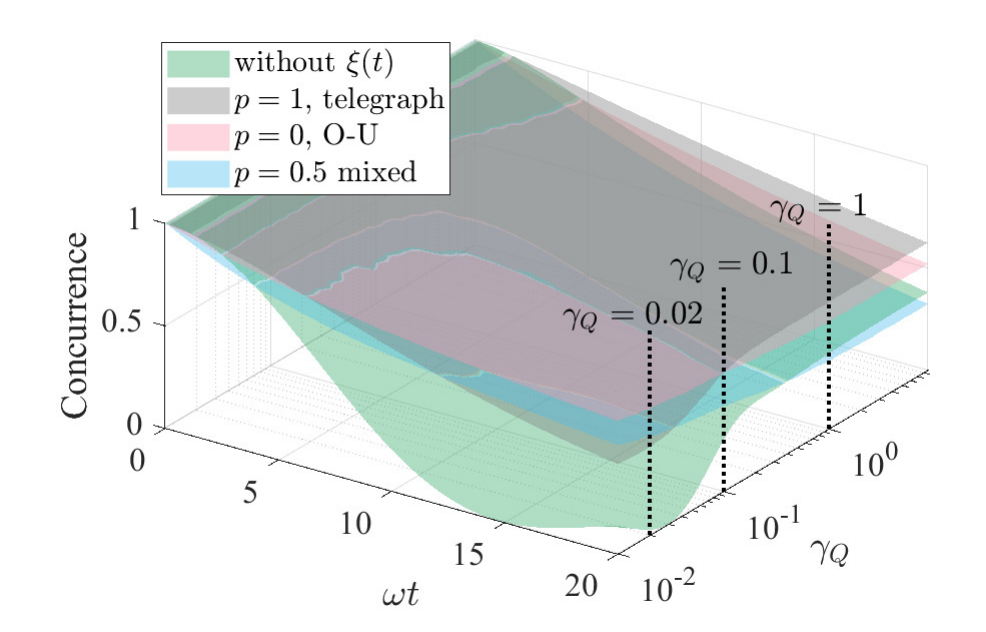}

\caption{\label{fig:4}Quantum entanglement protection induced by the mixture
of telegraph noise ($\xi_{a}$) and O-U noise ($\xi_{b}$). The gray
surface corresponds to telegraph noise only, the pink surface corresponds
to O-U noise only, and the blue surface represents the mixed noise.
As a comparison, the light-green surface shows the case without classical
noise. The parameters are $p_{{\rm jump}}=0.35$ (for $\xi_{a}$),
$\gamma_{\xi}=15$ (for $\xi_{b}$).}
\end{figure}
\begin{figure}
\includegraphics[width=1\columnwidth]{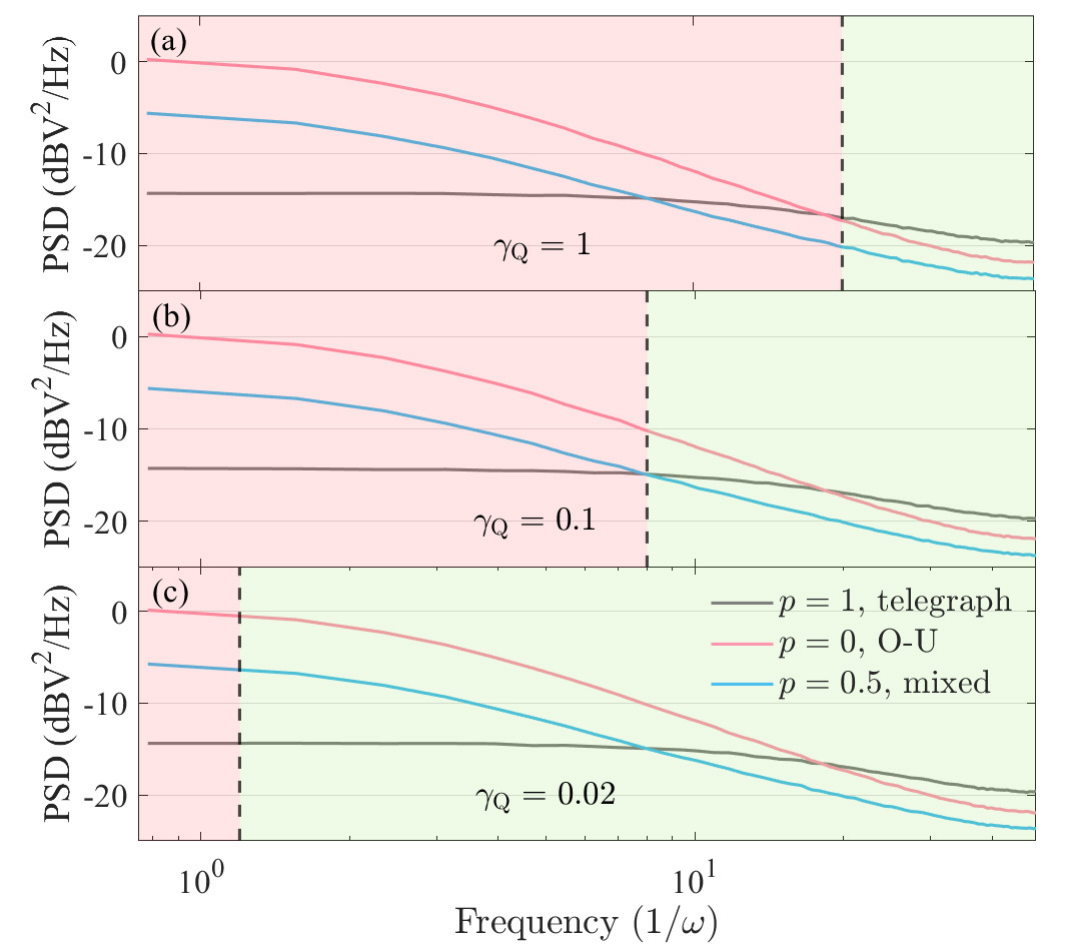}

\caption{\label{fig:5}The PSD of three types of noise are shown in (a), (b),
and (c), each corresponding to different values of $\gamma_{{\rm Q}}$
selected in Fig.~\ref{fig:4}. The gray, pink, and blue curves represent
telegraph noise, O-U noise, and mixed noise, respectively. The black
dashed line is a hypothetical boundary to distinguish LF and HF domains.
Since $\gamma_{{\rm Q}}$ are different in (a), (b), and (c), the reference dashed lines are also moved. This is different from Fig.~\ref{fig:3}.}
\end{figure}

In the previous subsection, we have demonstrated that HF classical
noises can suppress the dissipation caused by LF quantum noise. However,
the HF and LF mentioned here are relative. Specifically, the so-called
``HF'' refers to frequencies that are significantly higher than
the typical frequency of quantum noise. Therefore, the properties
of the quantum noise itself is also crucial to the entanglement protection.
In this subsection, we will show the impact of the quantum noise.
The mixed noise is chosen as a mixture of telegraph noise ($\xi_{a}$)
and O-U noise ($\xi_{b}$).

In Fig.~\ref{fig:4}, it can be observed that the performance of
entanglement protection is highly dependent on the intrinsic properties
of quantum noise. Here the property of the quantum noise is mainly
reflected by the parameter $\gamma_{{\rm Q}}$. From the PSD $J_{{\rm Q}}(\omega)$
in Eq.~(\ref{eq:JQ}), $\gamma_{{\rm Q}}$ determines the frequency
distribution of the quantum noise. A larger $\gamma_{{\rm Q}}$ implies
more HF components in the quantum noise. When $\gamma_{{\rm Q}}$
varies, the performance of entanglement protection for three types
of noises are also influenced. Here, we focus on three typical values
of $\gamma_{{\rm Q}}$ labeled in Fig.~\ref{fig:4}, $\gamma_{{\rm Q}}=1$,
$\gamma_{{\rm Q}}=0.1$, and $\gamma_{{\rm Q}}=0.02$. Similar to
the discussion in Sec.~\ref{sec:4A}, all the numerical can be explained
from the PSD plots in Fig.~\ref{fig:5}.

For the case $\gamma_{{\rm Q}}=1$ in Fig.~\ref{fig:5}(a), the order
of the three types of noise in terms of their proportion in the HF
part (light-green region) of the PSD, from highest to lowest, is telegraph
noise, O-U noise, and mixed noise. As can be seen from the numerical
results Fig.~\ref{fig:4}, the order of these three types of noise
in terms of the performance of entanglement protection is exactly
the same. This provides strong evidence that the ranking of their
HF proportion directly corresponds to the ranking of their entanglement
protection performance. Similar observations are in the cases $\gamma_{{\rm Q}}=0.1$
and $\gamma_{{\rm Q}}=0.02$. In Fig.~\ref{fig:5}(b) with $\gamma_{{\rm Q}}=0.1$,
telegraph noise and O-U noise exhibit nearly identical HF proportions,
and their performance of entanglement protection in Fig.~\ref{fig:4}
are consequently comparable. In Fig.~\ref{fig:5}(c) with $\gamma_{{\rm Q}}=0.02$,
O-U noise has the maximum HF proportion (best protection in Fig.~\ref{fig:4}),
and telegraph noise the minimum (worst protection in Fig.~\ref{fig:4}).

In conclusion, the proportion of the HF component is exactly positively
correlated with the performance of entanglement protection. Here,
we would like to point out the difference between the results in this
subsection and the previous subsection. In Sec.~\ref{sec:4A}, the
parameters of quantum noise are fixed, therefore we see the reference
line is always the same in Fig.~\ref{fig:3}. The difference in the
performance of entanglement protection is caused by the properties
of the classical noise. More precisely, $\gamma_{\xi}$ representing
the properties of classical noises determines the curves in Fig.~\ref{fig:3}.
In this subsection, the properties of quantum noise do not affect
the PSD of classical noises (curves in Fig.~\ref{fig:5}). But it
determines where are the reference lines as shown in Fig.~\ref{fig:5},
although the reference line is a phenomenological and approximate
description.

In summary, all numerical results in Fig.~\ref{fig:4} can be perfectly
explained using the data from Fig.~\ref{fig:5}. the above analysis
validates the role of quantum noise in the entanglement protection
mechanism. Specifically, our goal is to suppress the quantum noise
by higher frequency noises, then the properties of the quantum noise
defines what is a HF noise, thus affecting the performance of entanglement
protection for different classical noises.

\subsection{Influence of mixing ratios}

\begin{figure}
\includegraphics[width=1\columnwidth]{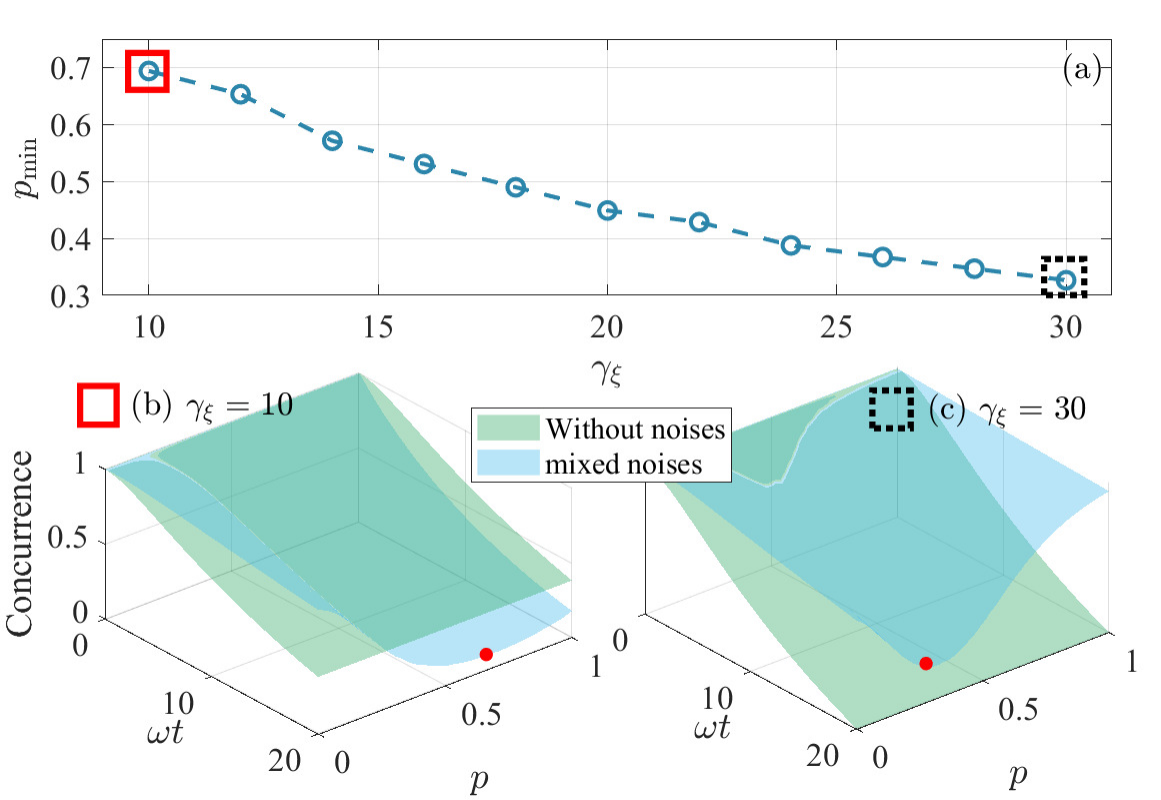}

\caption{\label{fig:6} Entanglement protection by the mixture of O-U noise
($\xi_{a}$, with mixing ratio $p$) and violet noise ($\xi_{b}$,
with mixing ratio $1-p$). (a) The mixing ratio for the minimum entanglement
at $\omega t=20$. Here, $p_{{\rm min}}$ denotes the mixing ratio
at which the minimum entanglement occurs {[}marked by red dots in
(b) and (c){]}. The detailed dynamical evolution of entanglement are
given in (b) and (c) for the case $\gamma_{\xi}=10$ (red solid square)
and the case $\gamma_{\xi}=30$ (black dotted square) respectively.}
\end{figure}
Additionally, we investigated the crucial significance of the mixing
ratio of the two types of noise for entanglement protection. In Fig.~\ref{fig:6},
the mixing ratio $p$ between O-U noise ($\xi_{a}$) and violet noise
($\xi_{b}$) is varying from 0 to 1. Then, we observe the mixing ratio
$p_{{\rm min}}$ that causes the minimum entanglement at $\omega t=20$.
In Fig.~\ref{fig:6}~(a), $\gamma_{\xi}$ increases from 10 to 30,
$p_{{\rm min}}$ shows a decreasing trend with increasing $\gamma_{\xi}$. 

This phenomenon can be interpreted as a competition between O-U noise
($\xi_{a}$) and violet noise ($\xi_{b}$). As we have emphasized,
better protection of quantum entanglement requires classical noise
to contain as many HF components as possible. In Fig.~\ref{fig:6},
the properties of the violet noise are fixed, while we change the
parameter $\gamma_{\xi}$ of the O-U noise. As $\gamma_{\xi}$ increases,
the O-U noise contains a higher proportion of HF components. Therefore,
increasing the weight of O-U noise (i.e., increasing $p$) in the
mixed-noise configuration is favorable to entanglement protection.
In other words, a moderate reduction in the weight of O-U noise (i.e.,
decreasing $p$) results in weaker entanglement, as illustrated in
Fig.~\ref{fig:6}.

In summary, the mixing ratio of different types of noises is crucial
to the performance of entanglement protection, and the selection of
the optimal ratio clearly depends on the properties of the constituent
noise components. Figure~\ref{fig:6} only presents a single illustrative
example; the scenario will become further complicated when the mixed
noise comprises a greater number of components.

\section{\label{sec:5}CONCLUSION}

We investigate the entanglement protection in a two-atom-cavity system.
Two types of noises are considered, the cavity leakage (modeled as
quantum noise) and stochastic atom-cavity couplings (modeled as classical
noise). Analytical derivations reveal the core mechanism: HF components
in the classical noise can freeze the initial quantum state of the
two atoms, thereby suppressing decoherence caused by the LF quantum
noise from cavity leakage and protect entanglement. 

Our numerical investigations further confirm that the performance
of entanglement protection by mixed noise is critically determined
by three key factors:

1. the intrinsic properties of the individual constituent noise components
and their specific mixing ratio, as these foundational elements collectively
shape the overall PSD of the mixed noise and lay the groundwork for
its entanglement protection capability;

2. the proportion of HF components in its PSD, which acts as the core
driver for suppressing decoherence induced by relatively LF noise
sources as it builds on the spectral features dictated by the constituent
noises and their mixing ratio;

3. the inherent characteristics of the quantum noise within the system,
which defines the relative boundary between HF and LF noise and thereby
modulates how effectively the HF components of the mixed noise can
counteract dissipative effects.

This work provides key insights for noise engineering in practical
quantum information processing, highlighting that optimizing the HF
content of mixed noise offers a viable strategy to mitigate decoherence
and protect entanglement in open quantum systems.

\bibliographystyle{apsrev4-2}
\bibliography{Ref6}

\end{document}